\documentstyle[12pt,aaspp4]{article}

\def\deg{^{\circ}}
\def\sun{\odot}
\def\inv{\ifmmode^{-1}\else$^{-1}$\fi}

\def\A{{\sf A}}
\def\B{{\sf B}}
\def\C{{\sf C}}
\def\X{{\sf D}}
\def\AB{{\sf A-B}}
\def\etal{{\it et~al.\ }}
\def\kms{km~s$^{-1}$}
\def\vsys{$v_{\rm sys}$}
\def\vlsr{v_{\rm LSR}}

\begin{document}
\title{Water Maser Emission and the Parsec-Scale Jet in NGC~3079}

\author{Adam S. Trotter, Lincoln J. Greenhill, James M. Moran \& Mark J.
Reid}
\affil{Harvard-Smithsonian
Center for Astrophysics, 60 Garden Street, Cambridge, MA 02138}

\author{Judith A. Irwin}
\affil{Queen's University, Stirling Hall, Rm 308E, Kingston, ON  K7L 3N6
Canada}

\author{Kwok-Yung Lo}
\affil{University of Illinois, 1002 W. Green St., Urbana, IL  61801}

\begin{center}
To appear in {\it The Astrophysical Journal}, Vol. 495, 10 March 1998
\end{center}

\begin{abstract}

We have conducted VLBI observations at sub-parsec resolution of
water maser and
radio continuum emission in the nucleus of the nearby active galaxy
NGC~3079.  The 22~GHz maser emission arises in
compact ($\sim 0.01$~pc at a distance of 16~Mpc) clumps,
distributed over $\sim 2$~pc along an axis that is
approximately aligned with the major axis of the galactic disk.
The Doppler velocities of the water maser clumps are consistent with their
lying in the inner parsec
of a molecular disk with a binding mass
$\sim 10^6 M_\sun$, rotating in the
same sense as the edge-on kpc-scale molecular disk observed in CO
emission.
However, the velocity field has a significant non-rotational
component, which may indicate supersonic turbulence in the disk.
This distribution is markedly different from that of water masers in
NGC~4258, which trace a nearly perfectly Keplerian rotating disk
with a binding mass
of $3.5\times 10^7 M_\sun$.
The 22~GHz radio continuum emission in NGC~3079 is dominated by a compact 
($< 0.1$~pc) source that is offset 0.5~pc to the west
of the brightest maser feature.  No bright maser emission is
coincident with a detected compact continuum source.  This
suggests that the large apparent luminosity of the maser is
not due to beamed amplification of high brightness temperature
continuum emission.  
At 8 and 5~GHz,
we confirm the presence of two compact continuum sources, with a
projected
separation of 1.5~pc.  Both have
inverted spectra between 5 and 8~GHz, and steep spectra
between 8 and 22~GHz.  
NGC~3079 may be a nearby, low-luminosity example of the class of
compact symmetric gigahertz-peaked spectrum (GPS) radio sources.
We detected a third continuum component that lies along the
same axis as the other two, 
strongly suggesting that this galaxy possesses a nuclear jet.
Faint maser emission was detected near this axis,
which may indicate a second population of masers
associated with the jet. 
\end{abstract}

\keywords{masers --- galaxies: active --- galaxies: nuclei 
--- galaxies: jets --- galaxies: individual (NGC~3079)}

\section{Introduction} \label{intro}

NGC~3079 is a nearly edge-on ($i=84\deg$) SBc galaxy at a distance
of $16$~Mpc ($H_0=75$~\kms~Mpc$^{-1}$; 1~mas corresponds to 0.08~pc).
Kiloparsec-scale radio lobes (\cite{dur88}) and
loops of H$\alpha$ emission (\cite{for86}) extend along the
minor axis of the galactic disk.    
Ionized material in these lobes shows evidence for a $\sim 2000$~\kms\
outflow (\cite{dur88}; \cite{fil92}; \cite{vei94}).
Heckman (1980) classified the optical emission spectrum of
NGC~3079 as that of a LINER, which could indicate shock-induced
ionization in the galaxy ({\it e.g.}, \cite{dop95}). 
Irwin \& Seaquist (1991) fit three-dimensional
galactic disk models to H~{\sc I} line emission, and determined
a systemic velocity of $1116\pm 1$~\kms\ with respect to the Local
Standard of Rest (LSR), for the radio astronomical definition of
velocity, $v/c = -\Delta\nu /\nu_0$ (at \vsys, $v_{\rm LSR, Rad}=
v_{\rm Hel, Opt}-0.9$~\kms). 
Observations of CO emission revealed
a dense, rotating molecular ring extending $\sim 200$ to 750~pc
from the nucleus along a position angle of $-15\deg$
(\cite{you88}; \cite{sof92}).  The disk rotates as a solid body
out to a radius of 400~pc, where the velocity is 330~\kms.
Within a 400~pc radius, the disk has a mean molecular density of
530~cm$^{-3}$ and a total molecular
mass $M_{\rm H_2} \sim 5\times 10^9 M_\sun$, 
or about 50\% of the dynamical mass in this inner region of the galaxy.
Irwin \& Sofue (1992) inferred a systemic velocity for the rotating CO
disk of $1145\pm 10$~\kms, nearly
40~\kms\ greater than the systemic velocity 
inferred for the galactic disk from H~{\sc i} emission (\cite{irw91});
the origin of this discrepancy is unclear.

Large-scale outflows of the type observed in NGC~3079 have been observed
in a number of galaxies that exhibit high far-infrared (FIR)
luminosities and evidence for starburst activity ({\it e.g.},
\cite{hec90}).  
However, Hawarden
{\it et al.} (1995) present a number of arguments against the starburst
hypothesis in NGC~3079:
\begin{itemize}
\item Despite an unusually large concentration of molecular gas in the inner
kiloparsec, the low ratio of 
FIR luminosity to molecular mass (\cite{you88}) indicates that the
star formation efficiency is not very high;  
\item The 10~$\mu$m emission does not appear to be centrally concentrated
(\cite{dev87}),
as would be expected if there were starburst activity in the nucleus;
\item The mid-IR spectral energy distribution is relatively flat, which
is atypical of starburst galaxies (\cite{haw86});
\item The inferred extinction towards the nucleus, $A_V \sim 7.5$~mag,
is lower than expected for a dusty starburst region.
\end{itemize}
An alternative model for the
origin of the outflow in NGC~3079 was proposed by Duric \& Seaquist (1988), who
suggested that a jet from a compact central engine could be directed along
the minor axis of the galactic disk by interaction with dense
gas in the nuclear environment.
This interaction should result in strong shocks in the molecular gas in
the nucleus.  The observed strength of
emission in the lower vibrational
transitions of H$_2$ is consistent with collisional excitation in strong
($v\sim 150$~\kms) shocks (\cite{haw95}).
Blue-shifted OH maser emission to the south of the nucleus (\cite{baa95})
may mark the boundary between a dense nuclear disk and a nuclear cavity swept
clear by outflowing material.

A 5~GHz VLBI observation of the nucleus of NGC~3079,
made in 1986 by Irwin \& Seaquist (1988), hereafter IS, revealed
two compact continuum components, each $\approx 12$~mJy and
with a projected separation of 1.5~pc
along an axis with a position angle $123\deg$ E of N.  Following the
convention
of IS, we will refer to the southeastern compact feature as component \A,
and the other as component \B.  
An extended feature (component \C) lying between \A\ and \B\ was
tentatively identified, but its characteristics were not
well-constrained by the observation.
The quoted 5~GHz flux density of the entire
nucleus (at sub-arcsecond resolution)
ranges between 60 and 100~mJy (see \cite{haw95}), which implies that
components \A\ and \B\ contribute at most 40\% of the total nuclear
continuum emission at this frequency.
IS argued that if the compact VLBI sources are features
in a jet confined by the ambient nuclear medium,
then the momentum flux in the inner parsec of NGC~3079 is
sufficient to power the kiloparsec-scale outflow (see also
\cite{dur88}).
Irwin \& Sofue (1992) find evidence in CO emission
of a component of molecular gas with high velocity dispersion, 
flowing away from the nucleus along the \AB\ axis.
They suggest that ``spurs'' in the CO images along a position angle 
of $\sim 30\deg$ may indicate a second outflow perpendicular
to the one along the \AB\ axis.

Pedlar \etal (1996) observed that the nuclear H~{\sc i} absorption profile
is comprised of a deep feature at 1140~\kms\ flanked by
red- and blue-shifted features offset at $\pm 100$~\kms. 
Baan \& Irwin (1995) observed
a similar profile in H~{\sc i} and OH absorption, 
although they found that the red-shifted absorption
peaks $\sim 20$~pc to the south of the other components.
VLBI images at 21~cm
(\cite{sat97}) showed an
H{\sc i} absorption feature at continuum component \B\ that is red-shifted
by $\sim 200$~\kms\ with respect to a similar feature at component \A;
absorption at velocities intermediate to these two features
is seen at some level against both components.

The nucleus of NGC~3079 is a site of intense H$_2$O maser
emission, the majority of which is distributed over a velocity range of
$\sim 100$~\kms, centered on
$v_{\rm LSR} \approx 1000$~\kms\ (\cite{hen84}; \cite{has85}).  
The flux density of the brightest feature, at 957~\kms, has varied
between
1 and 8~Jy over a 10-year
monitoring period (\cite{baa96}), 
with $\sim 4$~Jy variations occurring on timescales
of months. The average line width is $\sim 5$~\kms.
With the exceptions
of a 140~mJy feature at $v_{\rm LSR}=1190$~\kms\ (\cite{nak95})
and a faint feature at 1123~\kms\ (this paper), all of the known 
H$_2$O maser emission is blue-shifted relative to the systemic velocity.
It has been suggested ({\it e.g.} \cite{has90}; \cite{baa95})
that the extreme brightness of the H$_2$O maser in NGC~3079 could result
from highly beamed, unsaturated amplification of 22~GHz emission from
compact sources by foreground molecular gas.
A VLBI observation made in 1988 (\cite{gre90}) 
revealed that the majority of the maser emission is confined to a region
$\sim 0.2$~pc in diameter, distributed
among several compact clumps 0.002--0.02~pc in extent.  Two clumps 
with a projected separation of 0.08~pc lie
0.5~pc to the north of this region.  Several of the unresolved maser 
clumps exhibit emission over an unusually broad Doppler velocity range, up to
25~\kms. Multiple spectral features in the emission from a few of the clumps
indicate the presence of unresolved spatial components.
Velocity gradients were observed within some of the more extended clumps.

We present new images of the water maser emission and the 5, 8 \& 
22~GHz continuum emission in the inner parsec of
NGC~3079, made with the NRAO$^1$ Very Long Baseline Array 

\noindent
\hrulefill

\noindent
$^1$NRAO is operated
by Associated Universities, Inc., under cooperative agreement with
the National Science Foundation.

\noindent
\hrulefill

\noindent
(VLBA). Interpretation of the images is strengthened by the
determination of accurate relative
positions of the maser and the 22~GHz continuum
emission.
\S 2 describes our VLBA spectral line and continuum observations.
We present images of the water maser emission in \S 3, and images
of the 5, 8 \& 22~GHz continuum emission in \S 4.  \S 5 discusses the
implications of these results for models of the nucleus of
NGC~3079.  \S 6 summarizes our results.

\section{Observations} \label{obs}

We observed the $6_{16}-5_{23}$ transition of H$_2$O 
(rest frequency $\nu_0 = 22.23508$~GHz)
toward NGC~3079 for 8 hours on 1995 January 9, using
the NRAO VLBA and the VLA, which operated as a phased array of
27 antennas. 
The data were recorded in left circular polarization in 
four 8~MHz passbands of 512 spectral channels each
(0.21~\kms\ channel spacing).  The
band-center LSR velocities were 1150, 1020, 934 and 800~\kms,  
with the fourth bandpass tuned to lie outside of the
known velocity range of maser emission.
The data were processed on the VLBA correlator, and post-correlation
processing was performed with AIPS.  Amplitude
calibration included estimated corrections for
atmospheric opacity and elevation-dependent variations in antenna gain.
We calibrated residual delays and fringe rates, and complex bandpass
response with observations of 3C273, NRAO150 and 4C39.25.  
We referenced the phases of all spectral channels
to a single strong channel (at 954.5~\kms) on
the edge of the brightest maser feature.
To image the maser emission, we
coherently averaged the visibility data in sets of four spectral channels.
To image the continuum, we averaged over all the channels
in the bandpass tuned to lie outside the known range of maser emission.

The 5 and 8~GHz continuum observations were made on 1992 September 15 and 29.
The 5~GHz observations used an array consisting of the VLBA stations
at Hancock, Kitt Peak, and Los Alamos, the NRAO 140~ft antenna at
Green Bank, and the VLA phased-array.
The 8~GHz observations were made with a VLBI array consisting
of the VLBA stations at Hancock, Fort Davis, and Kitt Peak,
the NRAO 140~ft antenna, the 100~m antenna of the MPIfR
at Effelsberg, and the VLA phased-array.
The fringe calibrator sources for the continuum observations
were OJ287 and 0917+624.  
NGC~3079 was observed with a bandwidth of 28~MHz 
for a total of 2~hr at each frequency.
The data were correlated with
the MkIII correlator at Haystack Observatory.
Post-processing and synthesis imaging were performed in AIPS.
Table~\ref{tb-stat} lists the naturally weighted restoring beam and $1\sigma$ 
image noise for our observations.

\section{Water Maser Emission} \label{maser}

Figure~\ref{fg-bigmap} shows the angular and velocity distribution
of water maser emission in the nucleus of NGC~3079.
Figure~\ref{fg-spec} is a spectrum generated from
the total imaged flux density in each 0.8~\kms\ spectral channel.
The emission arises in compact ($\lesssim 0.02$~pc) ``clumps'', 
most of which lie in a 0.2~pc-diameter cluster
with a velocity range $930<\vlsr <1060$~\kms; this cluster includes
the maser peak (3.6~Jy at $\vlsr =957$~\kms), at the origin of the map in 
Figure~\ref{fg-bigmap}. Two additional clumps, 0.5~pc (7~mas) to the north,
overlap the main cluster in Doppler velocity, emitting
over a range $1000<\vlsr <1040$~\kms. The overall distribution of the
H$_2$O maser features we observed in 1995 January is similar to that seen in
1988 February (Greenhill 1990), although certain clumps have
faded while new ones have appeared.  
Clumps that appear to be common to both epochs coincide to $<0.1$~mas,
although proper motions between two epochs, for spectrally blended features,
cannot be estimated reliably.
The normalized fringe visibility amplitude of the
maser peak is $>0.9$ on our longest baseline (6500~km).
We therefore place an upper limit of 20~$\mu$as on the angular size of the
maser peak (linear size $\lesssim 5\times 10^{15}$~cm), 
and a lower limit on the peak brightness temperature,
$T_B > 3\times 10^{13}$~K.  Haschick {\it et al.} (1990) 
obtained a similar lower limit on the peak brightness temperature
with a four-station intercontinental
VLBI observation in 1986 October.
 
Some of the clumps are extended, and exhibit apparent 
gradients in radial velocity.
For example, the maser peak, at 957~\kms, lies in a $\sim 0.02$~pc
elongated clump (Figure~\ref{fg-bigmap}, {\it inset}) 
oriented roughly NW-SE, that exhibits a projected Doppler velocity
gradient of nearly 4000~\kms~pc$^{-1}$.  Gradients
of similar magnitude are present in other clumps, though
their position angles do not appear to
be correlated with each other or with any larger-scale nuclear structure.
If the masers do in fact lie in a nearly edge-on disk, extended clumps
are likely to be foreshortened, and so the apparent velocity gradients
should be regarded as upper limits.
Note that the blending of two or more features that are poorly
resolved in angle and velocity can mimic a velocity
gradient in an extended structure.  
Thus, the gradients we observe do not
necessarily imply the presence of coherent, rotating structures in the
molecular gas.  However, in the case of the extended clump containing
the maser peak, ``kinks'' in the structure belie the
presence of more than two unresolved sub-features, and these kinks at least
display a true dependence of Doppler velocity on position within
the clump, which cannot be attributed to spectral blending alone.

We detected two red-shifted maser features.  The first is a 
single 140~mJy feature, 1.2~pc SSE of the maser peak, at a
$\vlsr$ of 1190~\kms.  This feature was first identified in
a single-antenna spectrum of 1995 April (\cite{nak95}).
Including this red-shifted feature with the blue-shifted clumps
to the north, we find that the maser emission in the nucleus of NGC~3079 
extends a projected distance of 1.7~pc along an axis with a position
angle of about $-10\deg$ (Figure~\ref{fg-bigmap}).
This axis is approximately aligned with the major axis of the kpc-scale
galactic disk (position angle = $-15\deg$) 
as seen in optical (Nilson 1973), H~{\sc i} (\cite{irw91}) and CO (\cite{you88};
\cite{sof92}) images.  
The second red-shifted feature lies off of this north-south axis, 1.2~pc
SE of the 22~GHz continuum peak (\S\ref{contin}),
at a $\vlsr$ of 1123~\kms.  This feature is angularly coincident with
a faint, extended 22~GHz continuum source.  We note that a second maser
feature, at 1040~\kms, also lies off the north-south axis, 0.2~pc NW
of the 22~GHz continuum peak.  These off-axis features define a second
axis, with a position angle of $128\deg$, that is shared by the nuclear
continuum components. 
This suggests that there
is a second population of water masers 
in NGC~3079 that may be interacting with or amplifying the jet. 

\section{The Parsec-Scale Radio Continuum} \label{contin} 

Figures~\ref{fg-cmap}, \ref{fg-xmap} and \ref{fg-kmap}
show naturally-weighted 5, 8 and 22~GHz 
images of the parsec-scale continuum emission in the nucleus of NGC~3079.
Table~\ref{tb-contin} lists the fitted parameters of the compact continuum 
sources identified at each of these frequencies.
Figure~\ref{fg-sed} compares the flux densities
of the two brightest continuum components, \A\ and \B.

\subsection{5~GHz Continuum}\label{cband}

At 5~GHz we detected both continuum
components \A\ and \B\ of IS.  The angular separation of the peaks of
these two components is within 1.4~mas of that measured in 1986 
February, from which we place an upper limit on the relative
proper motion of the two continuum features of 
$200\ \mu$as~yr$^{-1}$ ($0.06c$).  Due to the limited $(u,v)$ coverage
of the 5~GHz VLBI observation, no substructure was resolved in either
component. Note that while the flux density of \B\ in 1992 is approximately the
same as that measured by IS in 1986, component \A\ appears to have
faded somewhat (Table~\ref{tb-contin}).
However, given the quoted uncertainties in the flux density measurements
of IS, it is not possible to determine with this observation whether
either component is in fact variable at 5~GHz.
We did not detect the extended continuum component \C\ reported by IS.  However,
component \A\ exhibits an extension to the W at the $15\sigma$ level
(Figure~\ref{fg-cmap}), 
and there is some indication of emission that is undersampled by our
shortest baselines.

We report the detection of a third continuum component
(labelled \X\ in Figure~\ref{fg-cmap} and Table~\ref{tb-contin}),
located $\sim 4$~pc SE of \B\ along the \AB\ axis.
Component \X\ is not visible at the
0.5 and 0.7~mJy/beam ($3\sigma$) level at 8 and 22~GHz, respectively.
Given the deconvolved size of this source at 5~GHz,
$\sim 4\times 3$~mas, it would be significantly resolved at 8 and 22~GHz 
(see Table~\ref{tb-stat}).
Experimentation with various subsets of the visibility data
indicates that component \X\ could have a flat spectrum and still be
undetectable in our higher-frequency images.  
However, we can rule out a rising spectrum above 5~GHz.  
This differs from \A\ and \B, both of which
clearly peak between 5 and 22~GHz (Table~\ref{tb-contin}).

\subsection{8~GHz Continuum}\label{xband}

At 8~GHz we also detected both
continuum components \A\ and \B.  Component \B\ is
comparatively strong, and extends $\sim 0.1$~pc 
along a position angle of $60\deg$.
Component \A\ is clearly composed of multiple features,
clustered within $\sim 0.2$~pc of each other.
The measured flux densities of these subcomponents range
from 2 to 7~mJy, and their angular sizes are typically $\lesssim 2$~mas
(0.15~pc).  Relative to \B\, the centroid position of component \A\
is $\sim 2$~mas to the SE (along a position angle
of $160\deg$) of that observed at 5~GHz.  This difference is likely due
to differences in the brightness distribution within \A\ at the two
frequencies; the 8~GHz substructure appears to 
trace the limb-brightened SE edge of
an extended source that is centered near the 5~GHz position.  The integrated 
flux densities of both \A\ and \B\ are greater at 8~GHz than at 5~GHz.
Assuming they did not vary significantly
in flux over the 12 days separating the observations,
we find both sources have inverted
spectra between 5 and 8~GHz: $\alpha_\A = -1.8$, and $\alpha_\B = -1.6$.

\subsection{22~GHz Continuum}\label{kband}

We imaged both continuum components \A\ and \B\ at 22~GHz 
(Figure~\ref{fg-kmap}) by phase-referencing the spectral visibilities
to the maser peak and averaging over an apparently line-free velocity
range of $750<\vlsr <850$~\kms.
The maser peak lies $6.7\pm 0.1$~mas to the ENE of component \B, at
a position angle of $81\deg$.  We detect no 22~GHz continuum emission
to a $3\sigma$ limit of 0.75~mJy/beam at the position of any of the
maser features in the north-south distribution described in \S\ref{maser}.
The 22~GHz emission is dominated by component \B\, with an integrated
flux density of $\sim 16$~mJy. 
Figure~\ref{fg-kmap}b was produced with a 100~M$\lambda$
Gaussian taper applied to the visibility weights in the $(u,v)$ plane,
in order to lower the angular resolution and thus reveal more extended
emission.  Inspection of the images
reveals that \B\ consists of a $\sim 10$~mJy unresolved
core, plus a $\sim 6$~mJy halo extending $\sim 0.1$~pc to the SSE.
Fainter halo emission is visible in the untapered image
at the $6\sigma$ ($\lesssim 2$~mJy/beam) level,
extending $\sim 0.2$~pc to the SE along the jet
axis.  (Figure~\ref{fg-kmap}a).
The integrated flux density of \B\
at 22~GHz in 1995 January is clearly less than it was at 8~GHz
in 1992 September (Figure~\ref{fg-sed}); if the component is not
intrinsically variable, we find a spectral index
$\alpha_\B=0.9$ between these two frequencies.  
However, the fitted size of this component at both frequencies is
$\lesssim 0.1$~pc, so flux variations on timescales of years 
would not violate light travel-time constraints.  

In the high-resolution 
22~GHz image (Figure~\ref{fg-kmap}a), continuum component
\A\ has a poorly defined morphology; it peaks 
at approximately the same position relative to
component \B\ as the easternmost compact subcomponent
observed at 8~GHz (Figure~\ref{fg-xmap}).  
Faint emission at the $3\sigma$ level extends 3~mas around the peak. 
In the lower-resolution tapered image (Figure~\ref{fg-kmap}b), component \A\ 
has an integrated flux density of $\sim 6$~mJy, suggesting that it
may have a steep spectrum between 8 and 22~GHz, $\alpha_\A \approx 1.0$.  
It is also possible that 
one or more of the compact subcomponents of \A\ is variable.

The low-resolution 22~GHz continuum image reveals an 
extended feature between components \A\ and \B, which we label \C.  
The centroid position of \C\ coincides within measurement errors to
the position of component {\sf C} seen by IS at 5~GHz.  
It is clearly elongated east-west over $\sim 7$~mas (0.5~pc). 
In the higher-resolution 22~GHz continuum image (Figure~\ref{fg-kmap}b), this
source appears
mottled at the $3\sigma$ level.  Images of intermediate angular resolution 
suggest that it consists of several compact
components distributed along an east-west axis.
We did not detect any emission between \A\ and \B\ in our 5 or 8~GHz images.  
If \C\ consists of multiple compact continuum sources, they must have
inverted spectra between 8 and 22~GHz.
Alternatively, it may be comprised of multiple, faint maser features
distributed between 750 and 800~\kms.
Note that the maser feature at $\vlsr =1123$~\kms\ is
coincident with the centroid of component \C\ (Figure~\ref{fg-bigmap}).
 
\section{Discussion} \label{discus}

\subsection{Compact Continuum Sources}\label{discuscont}

The detection at 5~GHz of three compact continuum features, separated from
each other by at least several parsecs and lying along a common axis, strongly
suggests that these sample a jet, 
rather than physically unconnected regions of starburst activity.
The question then arises as to which component (if any)
represents the central engine.
Baan \& Irwin (1995) suggest that component \A\ is the core and \B\ 
is part of the jet, based on the presence of H~{\sc i} and OH absorption features
at two distinct velocities in the nucleus.
(The angular separation of \A\ and \B\ is unresolved in their images.)  
If the absorption is due to material in a disk that rotates in the same
sense as the galaxy, the more blue-shifted absorption component should
be associated with continuum component \B, and that near the systemic
velocity with component \A, which they identify as the core.
However, recent high-resolution VLBI observations of H~{\sc i} absorption
(S. Satoh 1997, private communication)
indicate that the absorption profile towards continuum component \B\ is
red-shifted with respect to that towards component \A.
It is not clear whether the absorption profile 
reflects rotation, outflow
or more complicated motion in the absorbing gas, or whether
it is intrinsic to the compact sources.

Typically, the core component of a VLBI core-jet structure in a radio
galaxy is compact and has a flat spectrum.
In NGC~3079, continuum sources \A\ and \B\ both exhibit extended or
multi-component structure.  This is most apparent in the 8~GHz image
(Figure~\ref{fg-xmap}), where the morphology of
\A\ suggests a jet
lobe limb-brightened on its leading edge, while \B\ exhibits an extension
perpendicular to the jet axis.  Also, at 22~GHz (Figure~\ref{fg-kmap}), 
\B\ appears to consist of
an unresolved core containing 75\% of the total flux, plus a halo that
extends to the south and east.
Furthermore, neither component has a flat spectrum
(Figure~\ref{fg-sed}), but rather exhibit a turnover between 5 and 22~GHz
suggestive of free-free or synchrotron self-absorption.

The spectra of both sources are quite similar,
except that the flux density of \A\ is consistently lower than that of \B.
If one were a jet component and the other a compact core,
we might expect less similarity in their spectra.
The nucleus of NGC~3079 may be a low-luminosity, nearby example of
the class of compact symmetric objects (CSO's) that exhibit
gigahertz-peaked spectra (GPS) (\cite{phi82}; \cite{ode91}).  These sources
have been attributed to 
bipolar jets from AGN that are impeded
by unusually high-density nuclear gas.  In
the context of this model, the compact continuum components mark the
ram pressure--confined leading edges of a jet that is burrowing out through
the dense medium (\cite{rea96}).  The peaked continuum spectrum may
be due either to synchrotron self-absorption, or to free-free absorption
in ambient ionized nuclear gas.

It is likely that neither component \A\ nor \B\
marks the central engine of NGC~3079.  If the maser features
trace an edge-on disk that feeds the jet (\S\ref{discusmas}), 
the core should lie near the intersection of
the axes of the maser and continuum distributions.
Note also that \A\ and \B\ are roughly equidistant
from this point, as would be expected for
oppositely directed lobes of a non-relativistic jet.
We therefore suggest that
the central engine lies near the midpoint between
continuum components \A\ and \B.

Component \X\
was detected only at 5~GHz.  Non-detection at
8~GHz may be due in part to resolution effects.  However, an inverted
spectrum like that of \A\ and \B\ can be ruled out.  Since it
is nearly twice as far from the assumed position of the central engine,
component \X\ may have significantly different physical
characteristics than the other two. For example,
synchrotron losses may have altered the relativistic electron
energy distribution, or the source could have expanded as it travelled 
further along the jet.  If these three sources are synchrotron
self-absorbed emission, the spectrum of \X\ may turn over at
a lower frequency than that of \A\ or \B.
If extrinsic ({\it e.g.} free-free) 
absorption is also occuring, the different spectral
characteristics may be a function of different
lines-of-sight through foreground material.

Component \C, observed in the low-resolution 22~GHz image 
(Figure~\ref{fg-kmap}b), is close to, but 
not coincident with, the assumed position of the central engine.
Its position also agrees within the measurement errors with the
position of continuum component {\sf C} of IS; however, that source
was observed at 5~GHz, and was considerably larger (Table~\ref{tb-contin}),
so it is not clear that they are the same.
Component \C\ is unusual in that it was only detected at 22~GHz, which
suggests an inverted spectrum extending to high frequencies.  
It is possible
that \C\ is spectral in nature.  A cluster of
faint, blue-shifted maser features distributed over $\sim 100$~\kms\ in
Doppler velocity would
appear as a continuum source in the 22~GHz images, which were produced by 
averaging in frequency over an apparently line-free region of the spectrum
(\S\ref{obs}).
This hypothetical maser could amplify an undetected background continuum
source in the vicinity of the central engine.

\subsection{The Nature of the Maser Emission} \label{discusmas}
  
The sense of rotation of the galactic disk of NGC~3079 is such that
the northern side is approaching and the southern side is receding. (The
galaxy's inclination of $84\deg$ causes the near side of the disk to be
projected to the west of the major axis.)   The overall distribution of
maser Doppler velocities is roughly consistent
with their arising in the inner parsec of an edge-on molecular disk that
is rotating in the same sense as the kpc-scale molecular disk observed in CO
emission (Figure~\ref{fg-model}).  
The center of rotation would be approximately midway between
the main blue-shifted maser cluster and the 1190~\kms\ feature to its south.  
Note, however, that the velocity dispersion of the clumps
in the northern (blue-shifted) half of this maser
structure is comparable to the apparent rotation
velocity, $\Delta v\sim v_{\rm rot}\sim 100$~\kms.  Given that the masers
do not trace a clear rotation curve, it is difficult to estimate the mass
distribution of the inner parsec of NGC~3079.  However, the overall
velocity distribution is consistent with the presence of a binding
mass of $\sim 10^6 M_\sun$ within 0.5~pc of the center of the putative disk.
This system is significantly less well-defined than that in the center
of NGC~4258, where water masers trace a thin, moderately warped, Keplerian
disk of inner radius 0.13~pc, outer radius 0.26~pc, 
and a central mass of $3.5\times 10^7 M_\sun$
(\cite{miy95}; \cite{her97b}).  

If the masers in NGC~3079 emit isotropically,
the total luminosity in the 22~GHz transition
is $\sim 200 L_\sun$, in a volume $\lesssim 0.01$~pc$^3$. 
(For comparison, the brightest Galactic water maser source, W49N, has an
isotropic luminosity of only $\sim 1 L_\sun$ in a comparable volume.)
It is difficult to account for such a high luminosity with standard
maser pumping models (\cite{rei88}; \cite{eli92a}).
However, the true luminosity
is probably less than the isotropic luminosity by a factor $\Omega_b /4\pi$,
where $\Omega_b$ is solid angle into which the maser emission is beamed.
Maser amplification of
radiation from a compact background source naturally results in
small $\Omega_b$.  Earlier VLBI and VLA
observations established that the centroid of the
masing region in NGC~3079 was within 5~mas of the centroid of the nuclear radio
continuum source (\cite{has90}).  It was
predicted that at higher resolution, individual bright water maser 
features would be
seen projected against one or both of the compact continuum components
identified in the 5~GHz VLBI image of IS.
We have found, however, that no bright maser emission in this galaxy
is angularly coincident with any detectable compact continuum emission.
The main cluster of maser features is offset by $\approx 5$~mas (0.4~pc)
from the 22~GHz compact continuum peak (source \B).

Highly beamed emission is probably still occurring in NGC~3079.
First, the maser features may amplify compact continuum sources
that are below our detection threshold, although the observed
distribution of maser features would require a complex continuum morphology.
Second, superposition of two or more
maser features with similar radial velocities along the line-of-sight
can result in highly beamed emission (\cite{deg89}), although this model
may have difficulty reproducing the observed velocity range of bright
maser features, and the extended structures with velocity gradients.  
Third, beaming may result from exceptionally high aspect ratios in the
molecular gas.  For a saturated, 
velocity-coherent cylinder with a (length/diameter) aspect
ratio $a$, $\Omega_b \approx 1/a^2$ (\cite{gol72}; \cite{eli91}).

One model for the pumping of water masers in AGN relies
on irradiation of molecular gas
by X-rays from the central engine (\cite{neu94}).  
A dense cloud that is illuminated by
an external X-ray source will develop a warm layer in which the water abundance
is enhanced, and in which collisional pumping maintains a population inversion.
The presence of dust can greatly increase the depth of this layer
by absorbing the infrared photons produced in the pumping process that
would otherwise thermalize the molecular energy levels and quench the maser
(\cite{col95}; \cite{wal97}). 
However, the luminosity of the nucleus of NGC~3079 in soft (0.5--3~keV)
X-rays is
$\sim 1\times 10^{40}$~erg~s\inv\ (\cite{fab82}),
and there is no convincing evidence of a nuclear hard X-ray source, obscured
or otherwise, from
ASCA observations (K. Weaver 1997, private communication).
Hence, X-ray heating is probably not a significant source of pumping 
for the water maser in this galaxy.

An alternative pumping model invokes molecular shocks.
A fast ($\sim 100$~\kms), dissociative shock propagating through
a medium with a preshock density of
$\sim 10^7$~cm$^{-3}$ will leave a large column of warm
($\sim 400$~K) molecular gas in its
wake, in which the H$_2$O abundance is enhanced and collisional excitation
maintains the maser population inversion (\cite{eli89}; \cite{eli92b}).
The resulting sheet-like masing region, when viewed edge-on, provides
the long amplification pathlengths and the beaming geometry necessary to
produce high-brightness temperature maser features.
Slower, non-dissociative C-type shocks may also produce conditions
conducive to water maser emission (\cite{kau96}), and may be necessary
to explain the observed strength of higher energy level water maser
transitions in star-forming regions in our Galaxy ({\it e.g.} \cite{mel93}).
The detection of strong H$_2$ emission (\cite{haw95}), high-velocity
outflows ({\it e.g.}, \cite{fil92}), and dense molecular gas (\cite{you88};
\cite{sof92}) all suggest that molecular
shocks are present in the nucleus of NGC~3079.

How might shocks be driven in NGC~3079?
We suggest a model in which the masers
lie in the inner parsec
of a dense molecular disk that is shocked by interaction with 
a loosely collimated nuclear wind.  
If we assume that the orientation of the inner parsec of the molecular
disk in NGC~3079 is the same as observed on larger scales, {\it i.e.} nearly
edge-on with a position angle of $\sim -15\deg$ (\cite{sof92}), 
then it is clear
that the nuclear jet, traced by the compact continuum sources,
is not perpendicular to the disk.
We suggest that the jet is surrounded by a loosely collimated outflow, or
``sheath'',
which disturbs regions of the disk lying under the jet,
driving turbulence and shocks that give rise to maser emission.  
Centrifugally-driven outflows from the innermost regions of
magnetized accretion disks may be
a common feature in AGN ({\it e.g.}, \cite{kon94}), and could be a source of
the hypothesized jet sheath.
Note that an outflow originating in the inner accretion disk
will not remain axisymmetric about the jet axis,
but be redirected along the minor axis of the galaxy,
since motion in the plane of the dense molecular disk 
will be impeded.  The redirected outflow will then feed
into the kpc-scale radio lobes observed to extend along the minor
axis of NGC~3079 (see, {\it e.g.}, \cite{dur88}; \cite{vei94}).
While certainly not the only conceivable interpretation, this model 
has the attractive quality of
reducing the complexity of the nucleus of NGC~3079 to two
relatively simple entities: 
a dense molecular disk, and a misaligned jet with an
associated sheath-like outflow.  

Irwin \& Sofue (1992) detected a kinematic component of CO gas that is 
distinct from the rotating molecular disk, 
extending along the jet axis defined by the compact radio continuum
sources. 
(Note that this component accounts for the apparent warp in the
disk reported by Sofue \& Irwin [1992].)
This indicates that molecular gas is somehow being entrained by the jet
outflow.  The emission we observe at 1123 and 1040~\kms\ along
the jet axis (Figure~\ref{fg-bigmap}) suggests that the entrainment
may begin within one parsec of the central engine, where the gas density
is highest. 
Other galaxies exhibit similar characteristics.
Broad linewidth water maser emission is observed extending
along the axis of a jet in the LINER 
NGC~1052 (\cite{dia97}), while 
in NGC~1068, maser emission marks the deflection of the nuclear
jet by a molecular cloud (\cite{gal96}).
Thus, it appears that the interaction of jets with ambient molecular gas
is responsible for at least some of the water maser emission seen
in certain AGN.  The presence of water maser emission along the
inner parsec of the jet in NGC~3079 lends further credence to the
hypothesis that this jet is interacting with the ambient molecular material
quite close to the central engine,
and may be responsible for exciting the brighter emission presumed to 
lie in the disk itself.

An AGN disk wind will be ionized, and so lines-of-sight passing through
the outflow will suffer free-free absorption.  If the molecular disk is
inclined so that the western side is the closer, the
orientation of the jet is such that maser emission from the northern, 
blue-shifted side of the disk will not be absorbed.  The red-shifted emission
on the south side, however, will have to traverse the ionized sheath
and may be heavily absorbed.  This could explain 
the asymmetry between the blue- and red-shifted emission in NGC~3079.
The presence of multiple continuum jet components may indicate that
the outflow occurs in intermittent bursts; the separation of 
blue-shifted maser emission
into two distinct clusters may be an
effect of two distinct ``gusts'' of wind from the inner accretion disk.
We note, however, that no continuum source analagous to component \X\
was detected to the NW of component \B, and that the multiple features
may trace standing shocks in the jet rather than intermittent outflow.

Although the relative distribution of red- and blue-shifted maser
emission in NGC~3079 is consistent with their lying in a roughly edge-on
rotating disk, the rotation curve is masked by the the large peculiar 
velocities of individual features.
Most notably, the large spread in Doppler
velocity ($\Delta v \sim 100$~\kms) of features
in the main cluster containing the maser peak suggests that they are
embedded in a region of supersonic turbulence.
Such a situation has also
been observed in the Galactic star-forming region W49N, where measurement
of proper motions of water maser spots has allowed the determination of
their true, three-dimensional spatial velocities (\cite{gwi92}).
Statistical analysis of these velocities and the angular distribution of
these spots indicates that they are embedded in 
supersonic turbulence, which may be caused by fluid
instabilities in the boundary layer between the ambient medium and
a bipolar outflow driven by a young stellar jet (\cite{gwi94}; \cite{mac94}).
A similar situation may be occuring in NGC~3079. Instabilities
in the boundary layer between a nuclear wind and the disk could
result in supersonic turbulence.  Dense clouds embedded in this turbulence
could then collide, producing shocks and maser
features over a wide range of radial velocities (\cite{tar86}).

Several authors have considered
the acceleration of dense maser clouds by a
high-velocity, low-density wind ({\it e.g.} \cite{str73}; \cite{nor79};
\cite{elm79}).  Rescaling the wind model predictions of 
Duric \& Seaquist (1988) for our adopted distance of 16~Mpc, we
estimate a mass-loss rate for the wind of approximately $8 M_\sun$~yr\inv,
and an outflow velocity of about 5000~\kms.  Taking a typical maser cloud
to have a diameter of 0.02~pc (a typical clump size
in NGC~3079), a molecular hydrogen number density
of $10^9$~cm$^{-3}$, and a distance from the nucleus of 0.5~pc, the
ram pressure from this wind is capable of accelerating the cloud to
a velocity of 50~\kms\ over a distance of 0.05~pc on a timescale of
$\sim 2000$~yr.  
However, it is not clear that non-gravitationally bound clouds can
survive being accelerated in this fashion (see, {\it e.g.}, \cite{har87},
and references therein).  
It has been proposed that the masers in some AGN could arise in
clouds lifted off the surface of an accretion disk
by a centrifugally-driven wind (\cite{kon94}; \cite{kar96}).
Magnetic pressure gradients (the so-called ``melon seed'' force) 
and radiation pressure on
embedded dust are other possible acceleration mechanisms.  
Alternatively, the clouds could
form in the outflow itself via hydrodynamical instabilities, 
eliminating the need for an acceleration mechanism.
A shell driven into the ambient medium 
can fragment into dense clumps via
Rayleigh-Taylor and thermal instabilities (\cite{har87}; \cite{war90}).

\section{Summary} \label{conc}

We have found no evidence that the
unusually bright H$_2$O maser emission in NGC~3079 is 
beamed amplification of high brightness temperature
22~GHz continuum emission; the 22~GHz continuum peak lies
$0.4$~pc east of the maser peak in projection.  
Both continuum components \A\ and \B\ appear to have
inverted spectra between 5 and 8~GHz, and,
assuming no intrinsic variability,
spectral turnovers between 8 and 22~GHz.
These spectra may indicate synchrotron self-absorption, or
free-free absorption in ionized ambient nuclear gas.  
This source may be a low-luminosity,
nearby example of the class of compact, symmetric, gigahertz-peaked 
spectrum (GPS) radio galaxies.

Component \A\ shows substructure at 8~GHz that is 
suggestive of a jet lobe.  Component
\B\ is somewhat more compact, but also exhibits structure at 8 and 22~GHz.
A third continuum component (\X) was identified at 5~GHz; it
lies 4~pc SE of \B\ along the \AB\ axis, appears to
have a flatter spectrum than \A\ or \B,
given its non-detection at 8~GHz.
The detection of a third component along the continuum axis lends support to
the idea that these represent features in a nuclear jet rather than,
for instance, physically unconnected starburst regions.
It appears that none of these three sources marks the nuclear engine
of NGC~3079, which may be heavily absorbed.  
A more extended source (\C) was detected at 22~GHz only; it may in fact
be weak water maser emission that is amplifying undetected background
continuum.

Water maser features in NGC~3079
trace a roughly north-south extended structure,
consistent with the emission 
arising in a nearly edge-on disk aligned
with the kiloparsec-scale molecular disk.  
The distribution of red- and blue-shifted maser
features is consistent with rotation in the same sense as the larger
disk, with a binding mass on the order of $10^6 M_\sun$.  
However, the velocity dispersion of clumps within the distribution,
and the suggestion of internal velocity gradients within the
clumps, indicate that non-rotational, possibly turbulent
motions are significant.  The masers may arise in shocks driven by
a wide opening-angle nuclear wind.
Two additional features were observed along the axis of the nuclear
jet, and may represent a second population of maser emission.

\clearpage
\begin{deluxetable}{crrrr}
\tablecolumns{5}
\tablewidth{0pc}
\tablecaption{Properties of Naturally-Weighted Images \label{tb-stat}}
\tablehead{
\colhead{} & \multicolumn{3}{c}{Synthesized Beam} & \colhead{} \\
\cline{2-4} \\
\colhead{$\nu$} & \colhead{Maj. Axis} & \colhead{Min. Axis} & 
\colhead{P.A.} & \colhead{$1\sigma$} \\
\colhead{(GHz)} & \colhead{(mas)} & \colhead{(mas)} & \colhead{(deg.)}
& \colhead{(mJy/beam)} \\
}
\startdata
5 & 8.4 & 2.5 & 3 & 0.11 \\
8 & 1.8 & 0.6 & --34 & 0.16 \\
22\tablenotemark{a} & 1.15 & 0.98 & 5 & 0.22 \\
22\tablenotemark{b} & 1.15 & 0.98 & 5 & 2.8 \\
22\tablenotemark{c} & 3.34 & 2.98 & 47 & 0.27 \\
\enddata
\tablenotetext{a}{Continuum (100~\kms\ average)}
\tablenotetext{b}{Spectral-line (0.8~\kms)}
\tablenotetext{c}{Continuum with 100~M$\lambda$ $(u,v)$-taper}
\end{deluxetable}
  
\begin{deluxetable}{crrrrrr}
\footnotesize
\tablewidth{0pt}
\tablecaption{Nuclear Continuum Components in NGC~3079 \label{tb-contin}}
\tablehead{
  \colhead{Feature} 
& \colhead{$r$ (mas)}
& \colhead{$\phi$ (deg)}
& \colhead{Maj. Ax. (mas)} 
& \colhead{Ax. Rat.}
& \colhead{Pos. Angle (deg)}
& \colhead{$S_0$ (mJy)}
} 
\startdata
\cutinhead{22 GHz (1995 January)}
\B & \nodata & \nodata & $1.2\pm 0.1$ & $1.3\pm 0.2$ & $170\pm 10$ &
$16\pm 1$
\nl 
\A & $24.1\pm 0.3$ & $127\pm 2$ & $\lesssim 3$ & \nodata & \nodata &
$6\pm 1$
\nl
\C & $15.4\pm 0.5$ & $126\pm 2$ & $7.4\pm 0.6$ & $2.2\pm 0.5$ 
& $90\pm 10$ & $11\pm 2$ \nl
\cutinhead{8 GHz (1992 September)\tablenotemark{a}}
\B & \nodata & \nodata & $1.3\pm 0.1$ & $1.4\pm 0.2$ & $57\pm 4$
& $37\pm 2$ \nl
\A 
& $22.7\pm 0.3$ & $127.7\pm 0.7$ & $4.0\pm 0.4$ & $3\pm 1$
& $134\pm 5$ & $15\pm 2$ \nl
\cutinhead{5 GHz (1992 September)}
\B & \nodata & \nodata & $<2.3$ & \nodata & \nodata & $16.8\pm 0.4$ \nl
\A & $21.3\pm 0.2$ & $124.9\pm 0.6$ & $< 4.5$ & \nodata & \nodata 
& $6.0\pm 0.5$ \nl
\X & $54.5\pm 0.4$ & $124.7\pm 0.4$ & $4\pm 1$ & $\sim 1$ & \nodata 
& $4.3\pm 0.5$ \nl
\cutinhead{5 GHz (1986 June; Irwin \& Seaquist (1988))}
\B & \nodata & \nodata & $< 3$ & \nodata  & \nodata
& $12\pm 7$ \nl
\A & $20.2\pm 0.3$ & $122.6\pm 0.3$ & $< 3$ & \nodata & \nodata 
& $12\pm 5$ \nl
\C & $17\pm 5$ & $130\pm 50$ & $< 20$ & \nodata & \nodata 
& $<45$ \nl
\enddata
\tablenotetext{a}{Image convolved to a resolution of 2~mas to smooth out
substructure}
\end{deluxetable}

\clearpage
\begin{center}
Figure Captions
\end{center}

\figcaption[fg-bigmap.eps]
{Map of the distribution of 22~GHz H$_2$O maser emission in the
inner parsec of NGC~3079, observed with the
VLBA in 1995 January.   The origin of the angular axes is at the
phase-reference maser feature, at $v_{\rm LSR}=956$~\kms.
Shaded circles mark the positions of maser features detected 
at the 5$\sigma$ level in 0.8~\kms-wide
spectral channel images. 
The diameter of the circle is proportional
to log(Flux Density).  
Grayscale indicates the LSR Doppler velocity of
the spectral channel; 
velocities of selected features are also
indicated on the map.
The galactic systemic velocities 
determined from H~{\sc i} and CO observations are 
marked with arrows on the grayscale bar.
Dashed lines
indicate the major axis of the kiloparsec-scale molecular
disk and the axis of the nuclear jet.
The expected position of the central engine, at the intersection
of these two axes, is marked with a diamond.
Ovals indicate the positions
and approximate angular sizes of the 22~GHz
continuum features \A, \B\ and \C.
{\it inset}: Expanded view of the maser feature distribution within
0.1~pc
of the maser reference feature.
\label{fg-bigmap}}

\figcaption[fg-spectrum.eps]
{Spectrum of total imaged maser flux density.
The Doppler velocity spacing of the spectral channels is 0.8~\kms.  
The LSR systemic velocity of 1116~\kms\ is indicated with an arrow.
\label{fg-spec}}

\figcaption[fg-cmap.eps]
{Naturally-weighted, self-calibrated image of 5~GHz nuclear
continuum, from a VLBI observation of 1992 September.  Image contours
are -1, 1, 2, 3, 4, 5, 6, 7, 8, 9, 10, 12, 14, 16, 18, 20, 25, 30, 35,
40, 45 and 50~$\times$~0.3~mJy/beam
($1\sigma = 0.11$~mJy/beam).  Emission peak is at 16~mJy/beam (Component
\B).
The axis of the maser feature distribution and the axis of the nuclear
jet
are indicated by solid lines.  Boxes mark the positions of selected
maser features.
\label{fg-cmap}}

\figcaption[fg-xmap.eps]
{Naturally-weighted, self-calibrated image of 8~GHz nuclear
continuum,
From a VLBI observation of 1992 September.  Image contours are at
-1, 1, 2, 3, 4, 5, 6, 7, 8, 9, 12, 14, 16, 18, 20, 25 and 
$30\times 0.5$~mJy/beam  ($1\sigma = 0.16$~mJy/beam).
The axis of the H$_2$O maser emission and the axis of the nuclear jet
are indicated by solid lines.  Boxes indicate positions of selected
maser features.
\label{fg-xmap}}

\figcaption[fg-kmap.eps]
{(a) Naturally-weighted 22~GHz continuum image 
created by phase-referencing spectral visibilities to the maser peak and
coherently averaging over a
velocity range of $750<v<850$~\kms.
The contours are in units of 0.7~mJy/beam ($1\sigma = 0.22~$mJy/beam).
Solid lines indicate the axis of the H$_2$O maser distribution and
the axis of the nuclear jet.  Boxes indicate the positions of selected
maser features.  
(b) Same as in (a), but with a 100~M$\lambda$ taper applied
to the $(u,v)$ data before imaging.  Image contours are in units of
1~mJy/beam ($1\sigma = 0.27$~mJy/beam).
\label{fg-kmap}}

\figcaption[fg-sed.eps]
{Spectra of the compact
continuum components \A\ and \B\ at 5, 8 and 22~GHz.  The spectral
indices of both components between 5 and 8~GHz and between 8 and 22~GHz
are indicated.
\label{fg-sed}}  

\figcaption[fg-model.eps]
{A model for the nuclear region of NGC~3079.  Here,
the maser emission arises in a nearly edge-on disk that is aligned
with and rotates in the same sense as the kpc-scale disk seen in CO
emission (Sofue \& Irwin 1992).
Contours show the 8~GHz continuum emission (Figure~4), and squares indicate
the positions of the brighter maser clumps.
The pair of cones illustrate the
approximate opening angle of the jet; note that the inclination of the
jet to the line-of-sight has not been determined.
The central engine is assumed to lie near where the major
axis of the maser distribution and the axis of the jet intersect.
The disk likely extends to a larger radius
than is shown in the schematic illustration, and its thickness is
unknown.
Arrows indicate the direction of the 
kpc-scale outflow or ``superwind'' observed to
extend along the minor axis of the galaxy.  Note that the jet is
misaligned with the axis of the superwind.
\label{fg-model}}

\clearpage
\begin{center}
\plotone{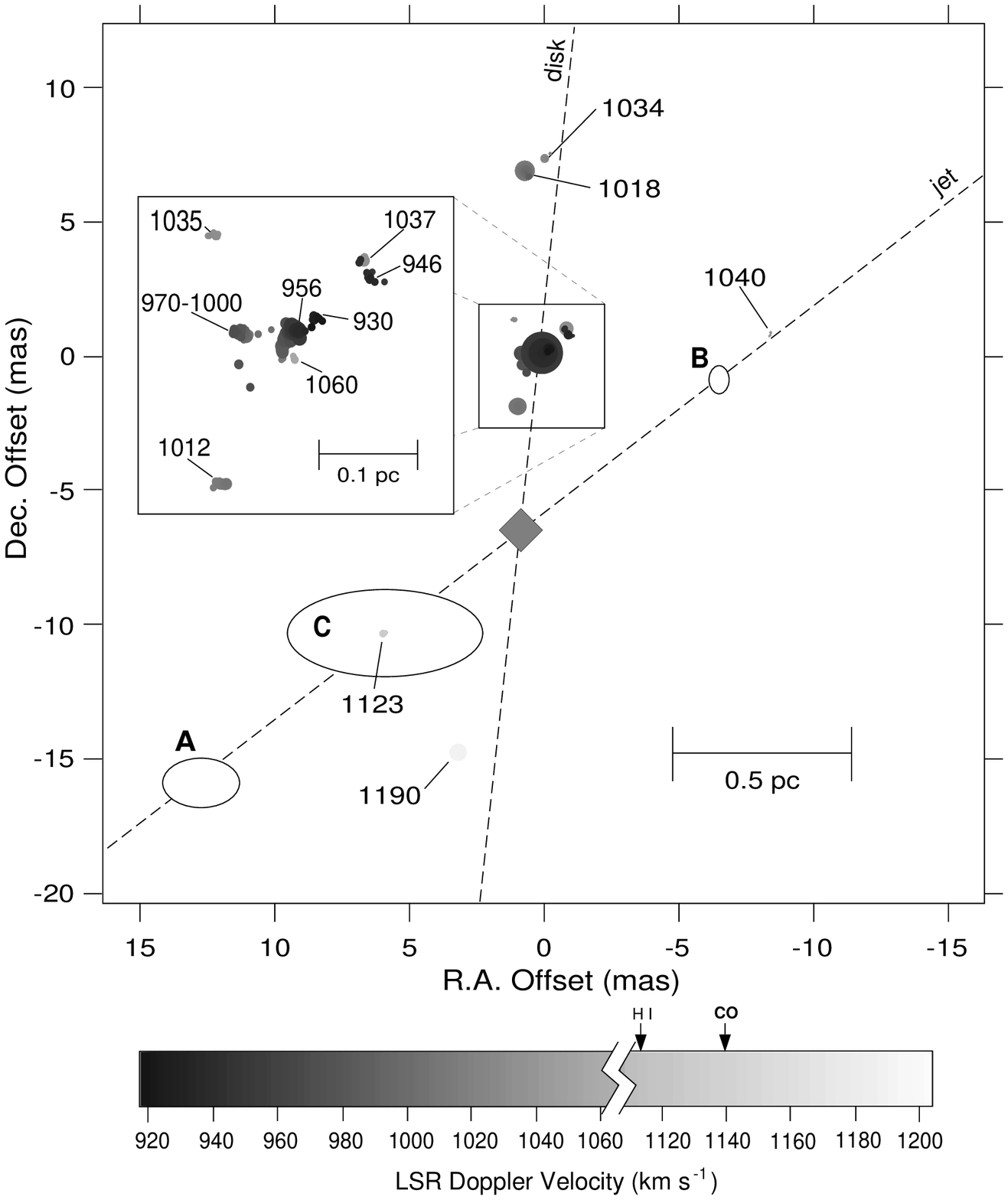}
\vfill
Trotter {\it et al.}, Figure 1
\end{center}

\clearpage
\begin{center}
\plotone{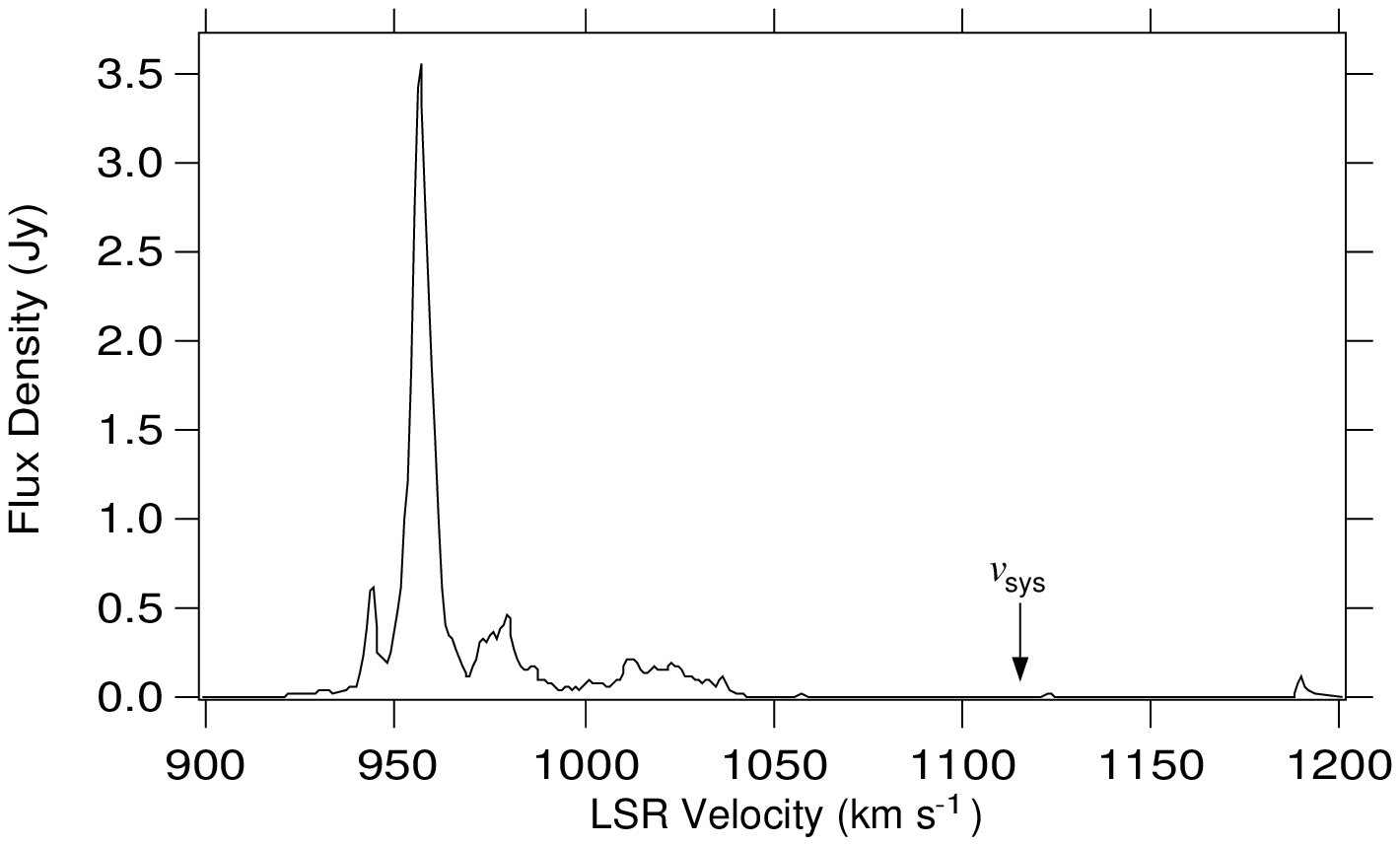}
\vfill
Trotter {\it et al.}, Figure 2
\end{center}

\clearpage
\begin{center}
\plotone{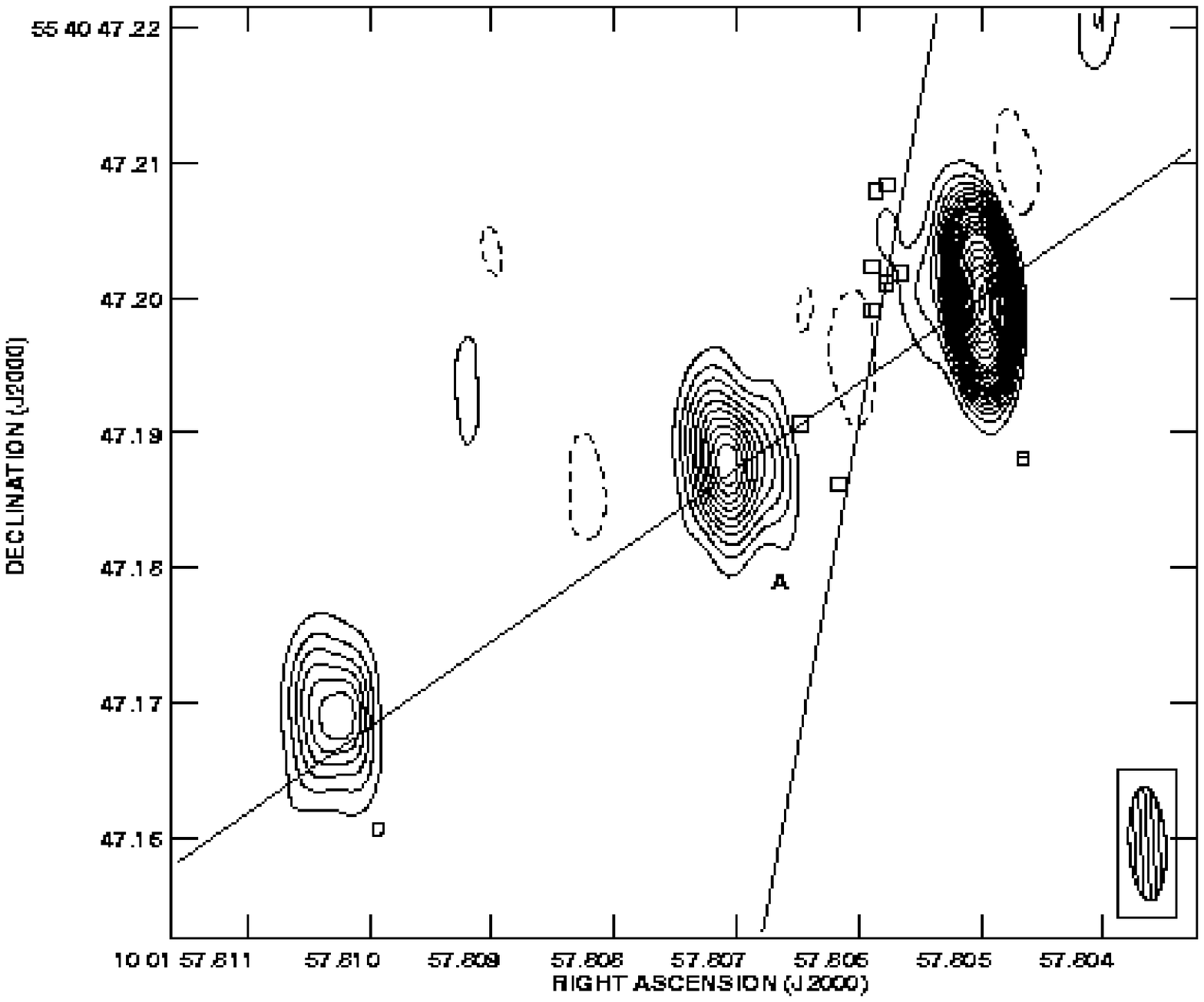}
\vfill
Trotter {\it et al.}, Figure 3
\end{center}

\clearpage
\begin{center}
\plotone{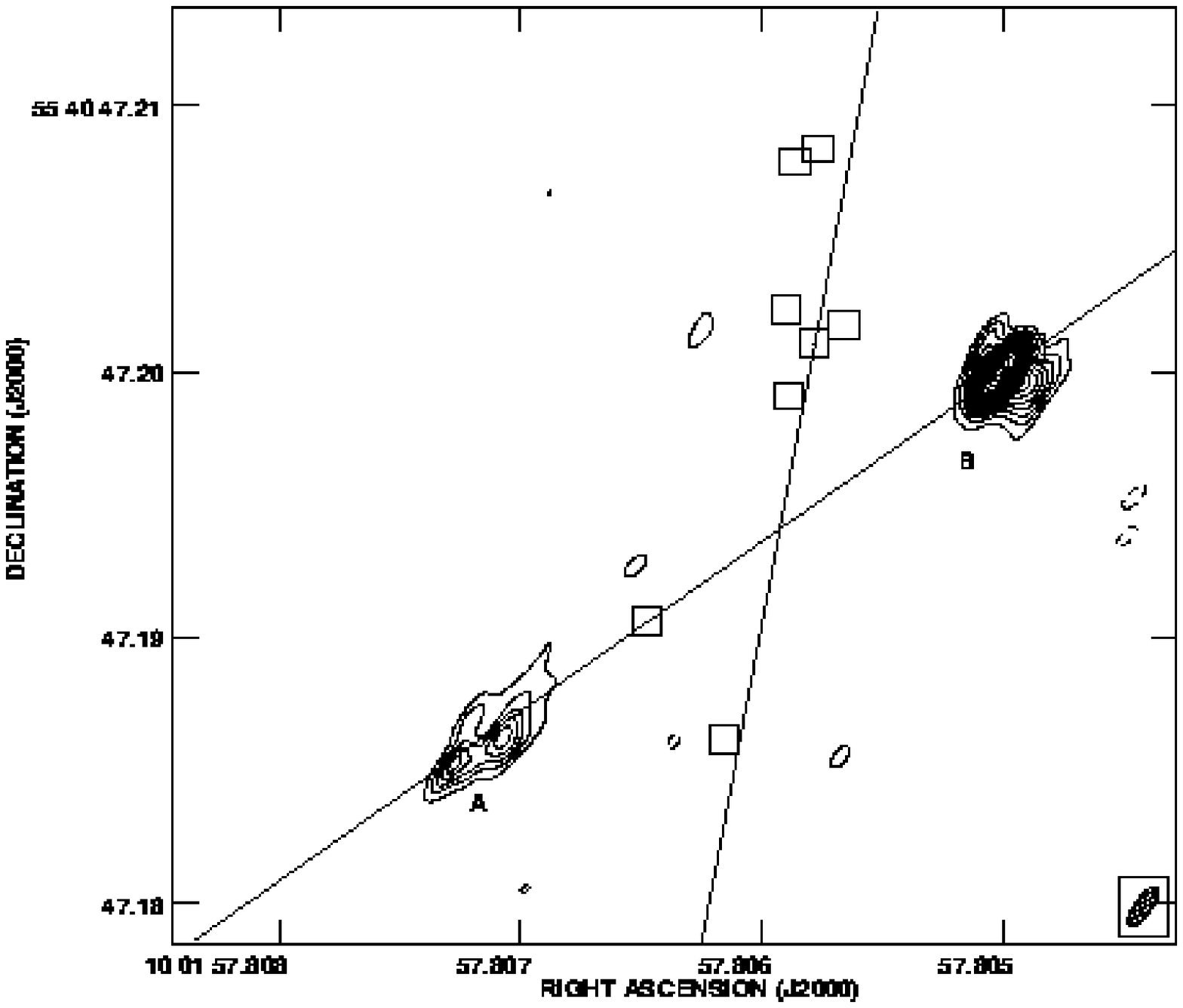}
\vfill
Trotter {\it et al.}, Figure 4
\end{center}

\clearpage
\begin{center}
\plotone{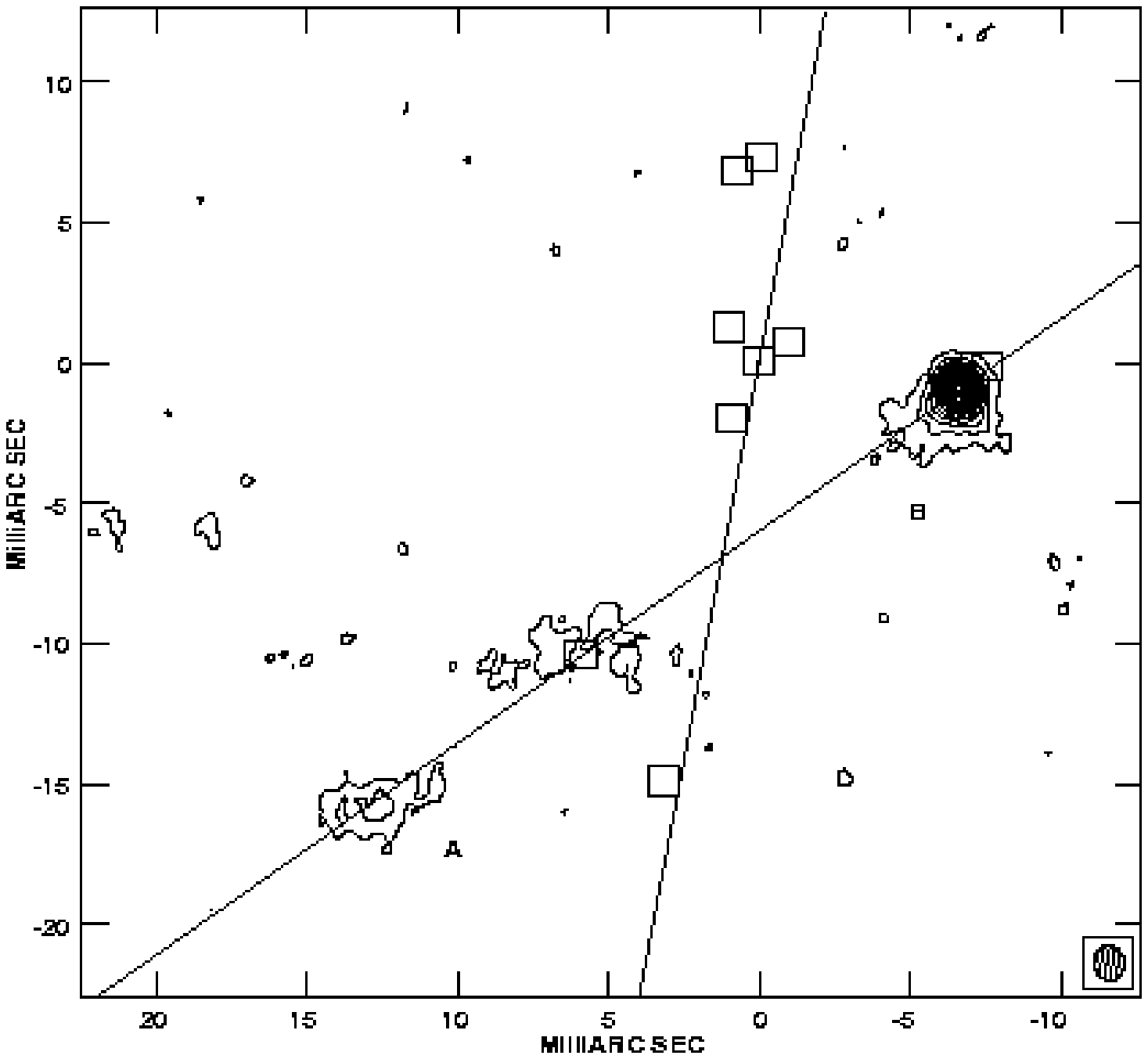}
\vfill
Trotter {\it et al.}, Figure 5a
\end{center}

\clearpage
\begin{center}
\plotone{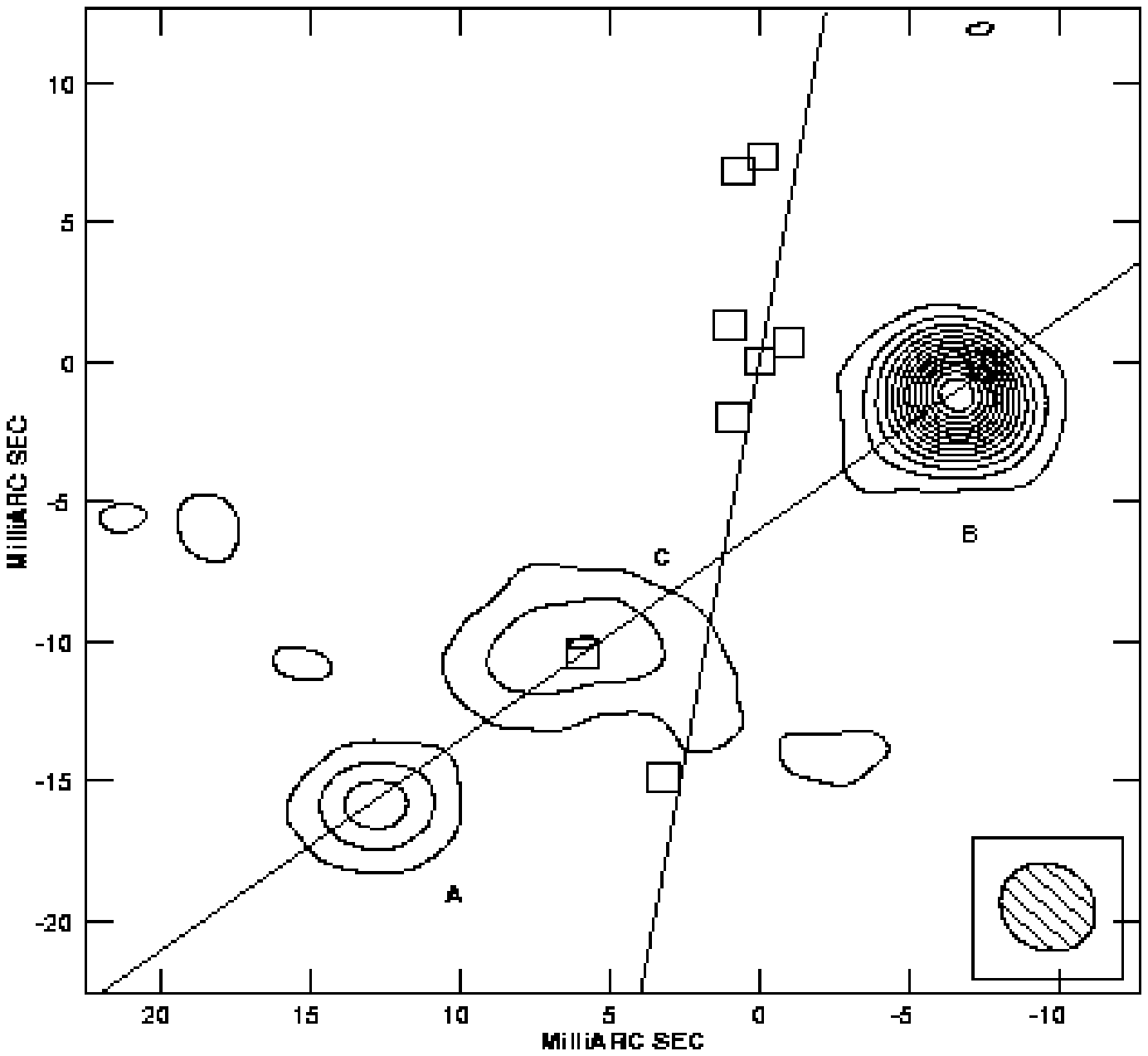}
\vfill
Trotter {\it et al.}, Figure 5b
\end{center}

\clearpage
\begin{center}
\plotone{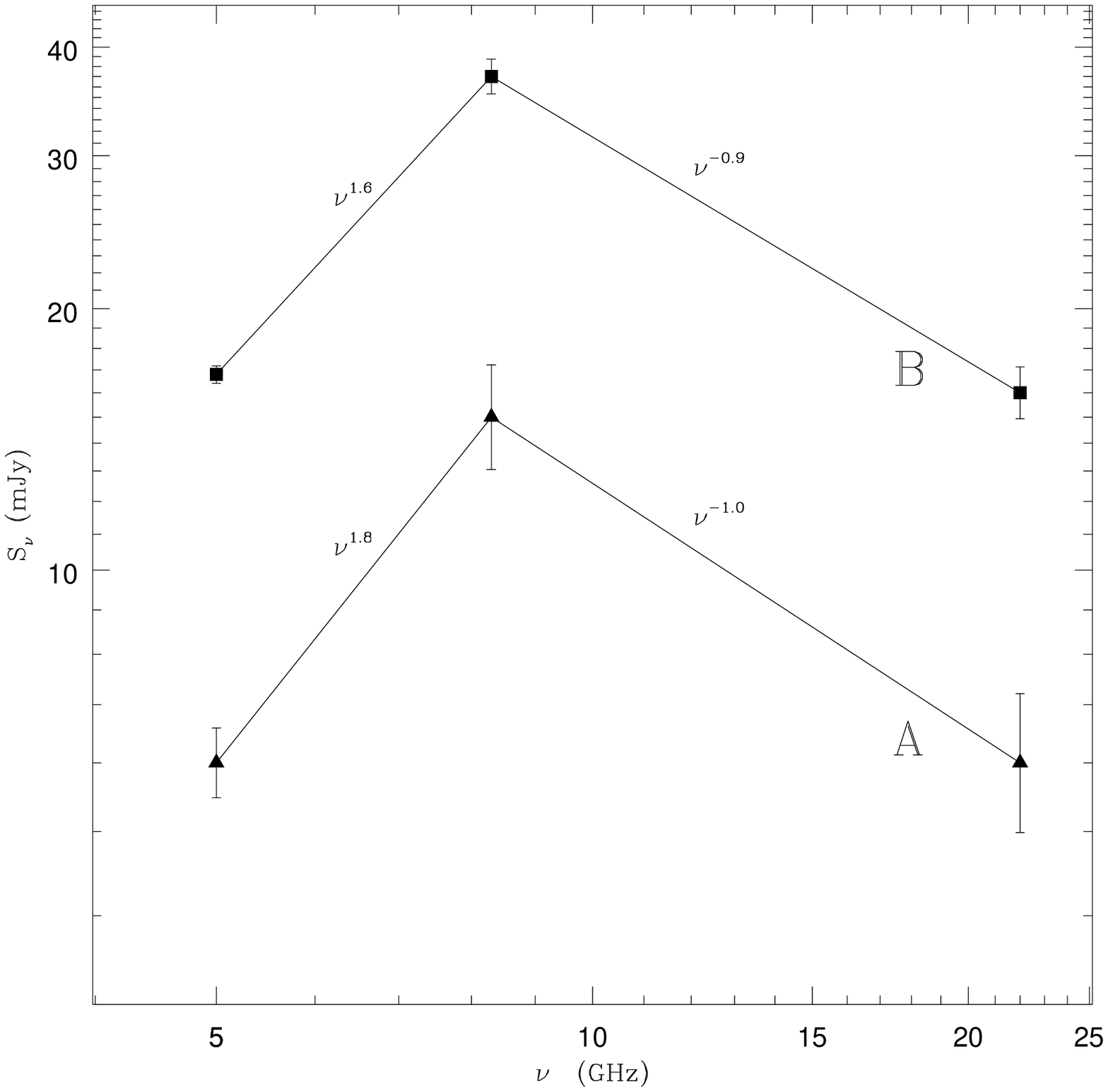}
\vfill
Trotter {\it et al.}, Figure 6
\end{center}

\clearpage
\begin{center}
\plotone{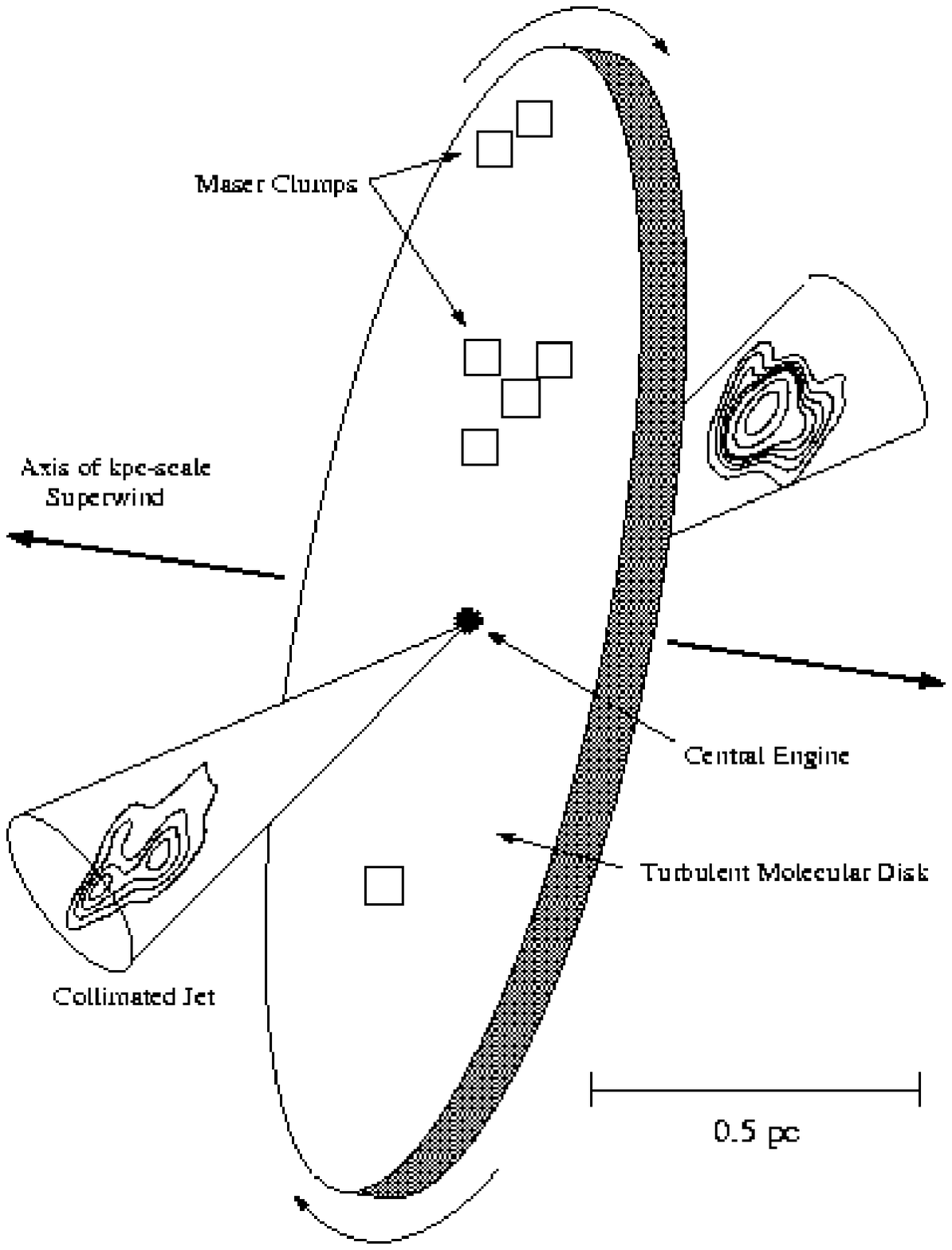}
\vfill
Trotter {\it et al.}, Figure 7
\end{center}

\end{document}